\documentclass[fp,twocolumn]{jpsj3}
\usepackage{txfonts}
\usepackage[version=3]{mhchem}
\usepackage{graphicx}
\usepackage{color}
\usepackage{tabularx}
\usepackage{array}
\usepackage{bm}
\usepackage{amsmath}
\usepackage{amssymb}
\usepackage{url}
\usepackage{braket}

\title{Polarized Neutron Diffraction Study on \ce{UPt2Si2}}

\author{Fusako~Kon$^1$, Chihiro~Tabata$^2$, Hiraku~Saito$^3$, Taro~Nakajima$^{3, 4}$\\ Hiroyuki~Hidaka$^1$, Tatsuya~Yanagisawa$^1$, and Hiroshi~Amitsuka$^1$}
\inst{$^1$Graduate School of Science, Hokkaido University, Sapporo, 060-0810, Japan\\
$^2$Materials Sciences Research Center, Japan Atomic Energy Agency, Tokai 319-1195, Japan\\
$^3$Institute for Solid State Physics, the University of Tokyo, Kashiwa 277-8581, Japan\\
$^4$RIKEN Center for Emergent Matter Science, Wako 351-0198, Japan}

\abst{We investigated the magnetic structure of the antiferromagnetic (AFM) ordered state ($T_{\rm N} \sim$ 34 K) in tetragonal \ce{UPt2Si2} using polarized and unpolarized neutron diffraction.
Previous neutron scattering studies reported that this system possesses a simple AFM structure with a propagation vector, ${\bm Q} = 0$, and the ordered magnetic moments aligned along the $c$-axis.
By contrast, our latest resonant X-ray scattering (RXS) experiments have revealed that the magnetic structure is modulated by the charge-density-wave (CDW) order, which emerges in one of the two Pt atomic layers in the unit cell below $\sim$ 320 K.
The modulation is characterized by a transverse wave in the $c$-plane, with the propagation vector of the CDW order, $\bm{q}_{\rm CDW } =$ ($\sim$0.42, 0, 0).
In the present study using neutron scattering, we observed that the superlattice reflections specified by $\bm{q}_{\rm CDW }$ develop below $T_{\rm N}$, in addition to the magnetic reflections with ${\bm Q} = 0$, thereby further confirming the presence of modulation in the AFM structure of this system.
From detailed analyses, we revealed that the amplitude of the transverse-wave magnetic modulation to be 0.72(2)~$\mu_{\rm B}$/U, which is a piece of quantitative information that could not be obtained through the RXS experiments.
This implies that the CDW drives the ordered magnetic moments to be tilted up to 20$^\circ$ in the AFM state. 
The observations strongly suggest that the magnetism of \ce{UPt2Si2} is heavily influenced by the hybridization effects between 5f electrons of U and 5d electrons of Pt.
}
\begin{document}
\maketitle
\section{Introduction}

The balance between kinetic energy, which gives mobility to electrons, and intra-atomic Coulomb interactions leading to electron localization, plays a crucial role in generating intriguing cooperative phenomena among electrons in solids, such as magnetism and unconventional superconductivity. 
This basic concept is further enriched by the orbital degrees of freedom of electrons in materials, which are influenced by crystal structures, local symmetries, and coupling with spins, enhancing the range of physical properties and driving the discovery of new phenomena in correlated electron systems\cite{Floquet_2005, Coleman_2007, Lohneysen_2007, Pfleiderer_2009, White_2015}. 

Particularly in U-based compounds, the distinct ``duality'' of itinerancy and localization of 5f electrons presents a notable challenge in the field of condensed matter physics. 
Classic but lasting examples include \ce{URu2Si2}\cite{Palstra_1985, Maple_1986, Schlabitz_1986, Mydosh_2020}, exhibiting an unknown ``hidden order'' coexisting with a heavy-electron superconducting state, and \ce{UPd2Al3} \cite{Geibel_1991, Sato_2001}, where unconventional superconductivity correlates with pronounced antiferromagnetic (AFM) ordering. 
The emergence of exotic superconductivity in \ce{UCoGe}\cite{Huy_2007}, \ce{URhGe}\cite{Aoki_2001} and \ce{UGe2}\cite{Huxley_2001} under ferromagnetic order, as well as in \ce{UTe2}\cite{Ran_2019} under a paramagnetic state with strongly anisotropic magnetic fluctuations, draws significant interest.
Moreover, \ce{URhSn} has recently revealed a novel hidden order coexisting with ferromagnetism \cite{Shimizu_2020}, which might involve quadrupolar ordering, sparking considerable intrigue.

Such fascinating properties of 5f electrons are believed to result from their spatial extent lying between that of more itinerant 3d and more localized 4f electrons. 
Due to this characteristic, whether to treat the 5f electrons as localized, perturbatively interacting with ligand ion orbitals like 4f electrons, or to start from a well-hybridized band incorporating electron correlations, remains unclear in many systems. 
Furthermore, in a U ion, the occupancy of two or three electrons in 5f orbitals necessitates a consideration of the balance between Hund's rule coupling and hybridization. 
It is theorized that due to this balance, 5f electrons might contribute to both heavy electron formation and localized magnetic responses (often 5f$^2$ $J =$ 4) across a wide temperature range\cite{Yotsuhashi_2001}. 
Additionally, strong spin-orbit coupling is believed to give rise to various higher-order multipolar degrees of freedom.

U$T_{2}X_{2}$ ($T$: Transition metal, $X$: Si/Ge) represents a family of systematically studied U-based intermetallic compounds. 
Most exhibit AFM ordering as their ground state, with variations in duality ranging from localized characteristics\cite{Amorese_2020} in \ce{UPd2Si2}\cite{Collins_1993, Honma_1998} to formation of spin density wave (SDW) reminiscent of itinerant magnetism in \ce{UCu2Si2} \cite{Honda_2006}. 
These systems are broadly understood within the framework of Doniach model \cite{Doniach_1977}, commonly applicable to 4f systems, where the competition between Ruderman-Kittel-Kasuya-Yosida (RKKY) interactions and Kondo effects due to d-f hybridization between $T$-ion d electrons and U 5f electrons dictates the variety in ground states. 
Yet, specific experimental data linking directly correlating d-f hybridization effects to the ground state characteristics of each compound are still scarce.

In this study, we focus on \ce{UPt2Si2} within the 1-2-2 series, aiming to unravel the correlation between $T$ and U ions. The Pt 5d electrons of this compound form a charge density wave (CDW) order, while U 5f electrons carry AFM order, coexisting at different ordering wave numbers. 
Thus, through diffraction experiments, we can directly investigate the interplay between these orders, with the expectation of enhancing our understanding of the d-f hybridization effects in U compounds.
\begin{figure}[b]
\begin{centering}
\includegraphics[width=0.95\linewidth]{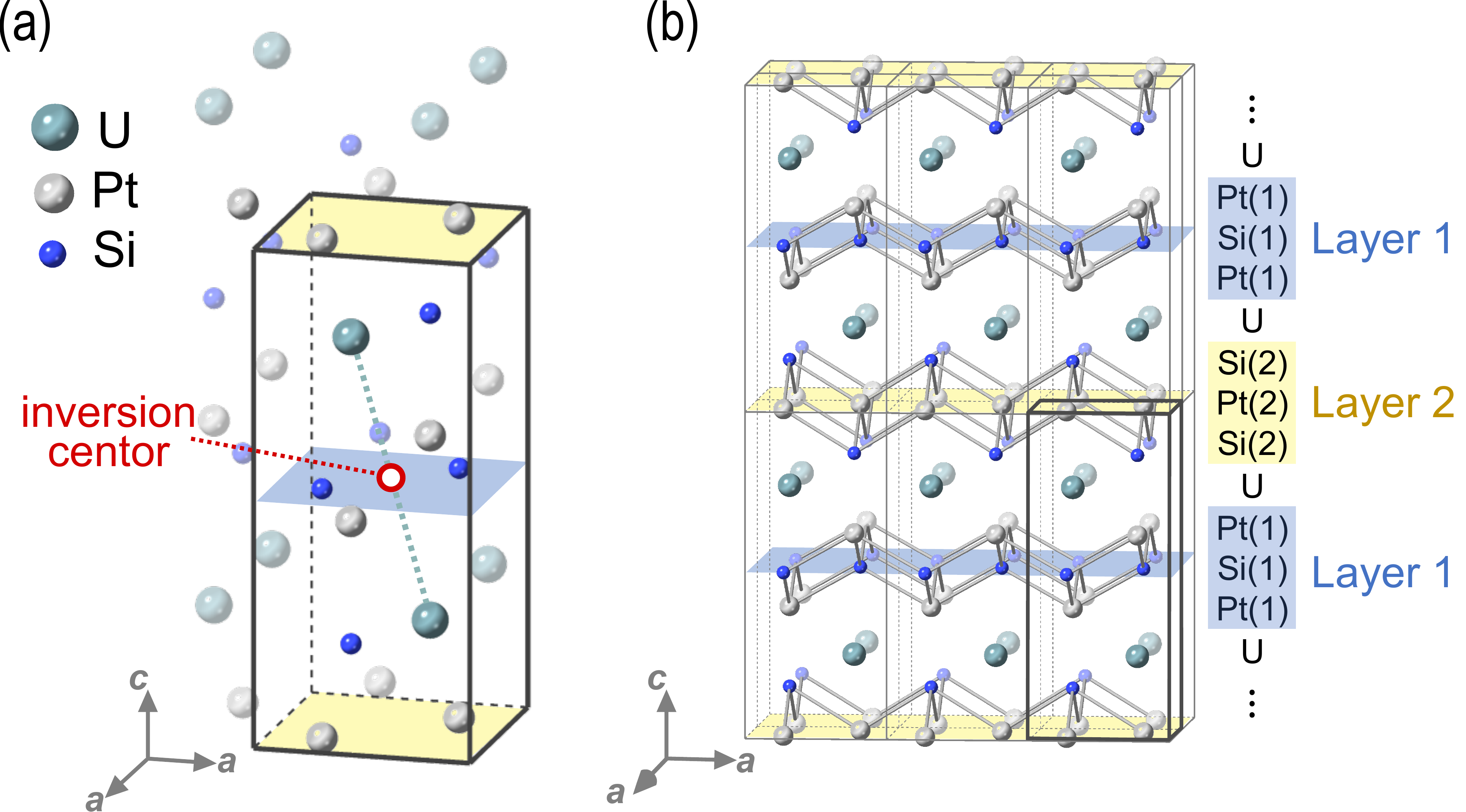}
\caption{(Color online) The crystal structure of \ce{UPt2Si2} shown from two perspectives: (a) the primitive unit cell and (b) the atomic layer structure. The bold lines outline the primitive unit cell and the blank red circle marks the inversion center. The colored shaded areas enhance the visibility of the atomic layers.}
\label{structure}
\end{centering}
\end{figure}

\section{Basic Properties of \ce{UPt2Si2}}
\ce{UPt2Si2} has a \ce{CaBe2Ge2}-type tetragonal crystal structure (space group: $P4/nmm$, No. 129, $D_{\rm 4h}^7$) \cite{Ptasiewicz_1985} (Fig. \ref{structure}), which exhibits a two-dimensional characteristic with atomic layers stacking perpendicularly to the $c$-axis.
Each U layer is sandwiched between a Pt-Si-Pt layer (hereafter referred to as Layer 1) and a Si-Pt-Si layer (Layer 2) alternately.
There is no spatial inversion symmetry at any atomic position, and the inversion center is located at the midpoint of the pairs of U atoms that span across either Layer 1 or Layer 2.

This system shows an AFM order below $34\pm1$ K ($\equiv T_{\rm N}$)\cite{Ptasiewicz_1985, Steeman_1990, Amitsuka_1992}.
Previous neutron scattering experiments have reported a magnetic structure in which the magnetic moments of U 5f electrons are aligned parallel to the $c$-axis with propagation vector of $\bm{Q} = 0$\cite{Steeman_1990, Ptasiewicz_1985, Sullow_2008}.
The estimated magnitude of the ordered magnetic moment varies across the references, ranging from 1.67 $\mu_{\rm B}/{\rm U}$\cite{Steeman_1990} to $2.5~\mu_{\rm B}/{\rm U}$ \cite{Sullow_2008}.
Given the large ordered magnetic moment and the fact that the electronic specific heat coefficient is $\sim$ 32 mJ/mol$\rm K^2$, indicating a slight increase in effective mass\cite{Amitsuka_1992}, the magnetism in this system is basically discussed in terms of a well-localized 5f-electron picture within the U$T_{2}X_{2}$ series.

Inelastic neutron scattering experiments\cite{Steeman_1988} have notably observed anomalies suggestive of crystalline-electronic-field (CEF) excitations, leading to the proposal of a singlet-singlet-doublet-singlet level scheme with $J = 4$ (5f$^2$)\cite{Nieuwenhuys_1987}.
Although this model qualitatively reproduces the magnetic anisotropy and the AFM ordering, the calculated ordered moment  (2.9 $\mu_{\rm B}/$U) significantly exceeds the experimental value. 
This discrepancy is speculated to result from the shielding of magnetism due to the Kondo effect, as discussed in many f-electron systems, through the d-f hybridization effects.
On the other hand, in recent inelastic neutron scattering experiments using a single crystal, coexistence of less-dispersive diffuse scattering with a resonant gap of $\sim$ 7 meV, characteristic of itinerant magnetism, and a substantial temperature-independent magnetic spectral weight (corresponding to $J =$ 4), typical of localized magnetism, along with transverse magnetic fluctuations have been observed \cite{Lee_2018}. 
This experimental fact highlights the dual nature of 5f magnetism as a challenging theme in the study of U-based compounds.

Meanwhile, recent elastic neutron and non-resonant X-ray scattering experiments\cite{Lee_2020}, as well as resistivity and specific heat measurements \cite{Bleckmann_2010}, have revealed that the system exhibits a CDW order by the 5d electrons of Pt with a propagation vector of $\bm{q}_{\rm CDW } =$ ($\sim$0.42, 0, 0) below $\sim$ 320 K ($\equiv T_{\rm CDW}$).
The formation of CDW is a characteristic commonly observed in Pt-based compounds with the \ce{CaBe2Ge2}-type structure, and it is believed to arise from the nesting instability of the Fermi surface, primarily contributed by the 5d electrons of Pt in Layer 2 \cite{Aoyama_2017, Kim_2015}.
In \ce{UPt2Si2}, the AFM order due to U 5f electrons coexists with this CDW below $T_{\rm N}$ \cite{Lee_2020}. 
Therefore, the study of the effect of the CDW on the magnetism is expected to give insight into the correlations between U 5f and Pt 5d electrons.
Diffraction experiments, which can separate CDW effects based on wave numbers, provide a means to gather microscopic insights into the correlation between 5f and 5d electrons in this system.

Our recent resonant X-ray scattering (RXS) experiments at U $M_{4}$ absorption edge revealed two distinct types of 5f electron resonance signals related to $\bm{q}_{\rm CDW }$ \cite{Kon_2023, Kon_2023_2}: a magnetic signal below $T_{\rm N}$ and a non-magnetic signal below $T_{\rm CDW}$. 
The first signal originates from a transverse magnetic modulation, where magnetic moments are tilted from the $c$-axis, generating a $c$-plane component ($c$-plane transverse magnetic modulation). 
The second signal stems from orbital modulation of 5f electrons, indicative of a quadrupole density wave. 
While our findings confirm the interplay between U 5f and Pt 5d electrons, the RXS experiments lacked comprehensive quantitative information. 
To address this, we performed polarized and unpolarized neutron diffraction measurements for an in-depth analysis of the magnetic modulation.

\section{Experimental Procedure}
The rod-shaped single-crystalline sample, grown along the $c$-axis, was synthesized using the Czochralski method.
No additional heat treatment was applied.
The obtained sample was evaluated by laboratory X-ray powder diffraction at the room temperature.
All observed peaks were indexed with the \ce{CaBe2Ge2}-type structure, and no impurity phases were detected within the S/N ratio of approximately 0.5\%.
Lattice constants were obtained as $a =$ 4.198(7) \AA~and $c =$ 9.69(2) \AA, consistent with those reported in the previous study\cite{Sullow_2008} within the range of experimental error.

\begin{figure}[h]
\begin{centering}
\includegraphics[width=0.95\linewidth]{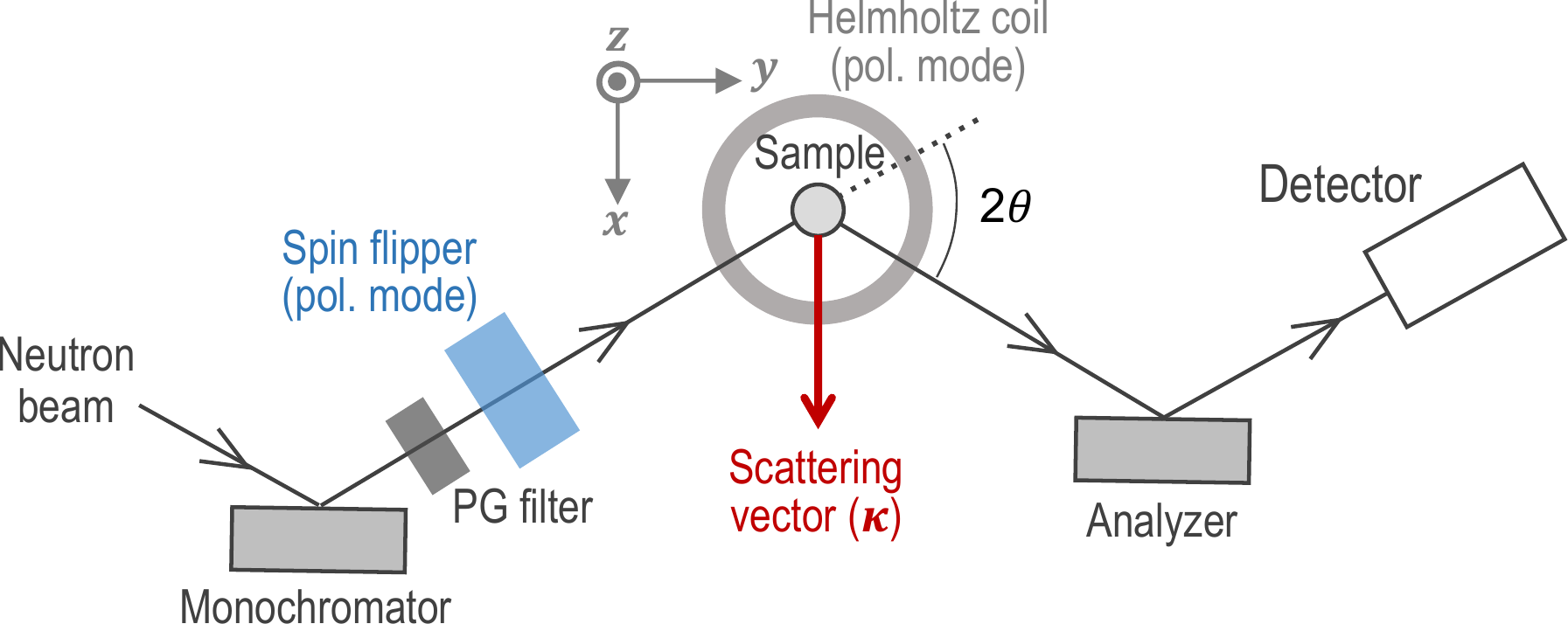}
\caption{(Color online) The schematic view of a triple-axis spectrometer (PONTA). The Cartesian coordinate $xyz$ is introduced with the $z$-axis perpendicular to the scattering plane and the $x$-axis parallel to the scattering vector $\boldsymbol{\kappa}$.}
\label{setting}
\end{centering}
\end{figure}

The neutron diffraction experiments were performed by using the polarized neutron triple-axis spectrometer PONTA in Japan Research Reactor 3 (JRR-3).
The schematic view of the spectrometer is illustrated in Fig. \ref{setting}.
The measurements were performed in the following two modes.
First, the ``unpolarized mode'' utilizes neutrons with the energy of $E =$ 34.06 meV ($\lambda =$ 1.55 $\AA$), without spin polarization pyrolytic graphite (PG) crystals served as both the monochromator and analyzer.
Second, ``polarized mode'' employs neutrons with $E =$ 14.7 meV ($\lambda =$ 2.36 $\AA$) monochromatized and polarized by a Heusler crystal.
Simultaneously, the spin state of scattered neutrons is analyzed in the polarization direction of the incident neutrons using another Heusler crystal.
In this mode, to determine the change in the neutron spin state before and after scattering, the spin polarity of the incident neutrons is controlled by turning on and off the spin flipper.
The device is positioned just before the sample, as illustrated in Fig. \ref{setting}.
When the flipper is activated, a spin flip (SF) signal is detected, indicating that the neutron spin flips from down to up during the scattering.
Conversely, when the flipper is deactivated, a non-spin flip (NSF) signal is observed, signifying that the spin state remains up throughout the scattering process.
The direction of spin polarization at the sample position is controlled by the magnetic field ($\sim$1 mT) generated by the Helmholtz coil.
In the present study, we used two configurations: the $P_{xx}$ setting, which polarizes the neutron spin parallel to the scattering vector, and the $P_{zz}$ setting, which polarizes the neutron spin perpendicular to the scattering plane.
Here, we adopt a coordinate system with the $x$-axis parallel to the scattering vector, and the $z$-axis perpendicular to the scattering vector, as depicted in Fig. \ref{setting}.

In polarized neutron diffraction measurements, three distinct signals can be differentiated: nuclear scattering, magnetic scattering from the $y$ component of magnetic moment ($M_{y}$), and magnetic scattering from the $z$ component of magnetic moment ($M_{z}$).
This distinction can be achieved by analyzing the SF and NSF signals in the two aforementioned settings\cite{Squires_2012}.
The relationships between the SF/NSF signals and their respective scatterers in the $P_{xx}$ and $P_{zz}$ settings are concisely summarized in Table \ref{polmode}. 
In the present experimental setup, the spin polarizations of the neutron beam were estimated to be $P_{0} =$ 0.8842 in the $P_{xx}$ setting and $P_{0} =$ 0. 9116 (in $P_{zz}$ setting). These values were measured using (2, 0, 0) nuclear Bragg reflection of the sample.
We used a $^{3}$He 0-dimensional detector in both modes.
The temperature at the sample position was controlled by a closed-cycle GM (Gifford-McMahon) refrigerator, with a range spanning from 2.3 K to 50 K.
Contamination of neutrons with higher order wavelengths was reduced by the PG filter.
All the measurements were performed within the ($hk$0) scattering plane.

\begin{table}[h]
\caption{The relationship between signal types and their origins in polarized neutron diffraction measurements with $P_{xx}$ or $P_{zz}$ setting. Non-spin-flip (NSF) scattering refers to the case where the spin of the scattered neutron does not invert from the incident neutron spin state, while spin-flip (SF) scattering refers to the case where the spin of the scattered neutron inverts from the incident neutron spin scate.
${M}_{z}$ denotes the component of the magnetic moment that is perpendicular to the scattering plane, while ${M}_{y}$ denotes the in-plane component, which is perpendicular to $\boldsymbol{\kappa}$.}
\label{polmode}
\vspace{1\baselineskip}
\centering
\begin{tabular}{ccc}
\hline\hline
&$P_{xx}$& $P_{zz}$ \\ \hline
SF  & $M_{y}$, $M_{z}$ & $M_{y}$\\ \hline
NSF & nuclear & nuclear, $M_{z}$ \\ \hline
 &   &
\vspace{-1\baselineskip}
\end{tabular}

\end{table}
\section{Experimental Results}
\subsection{$\bm{Q} = 0$ magnetic reflections}
\begin{figure}[h]
\begin{centering}
\includegraphics[width=1\linewidth]{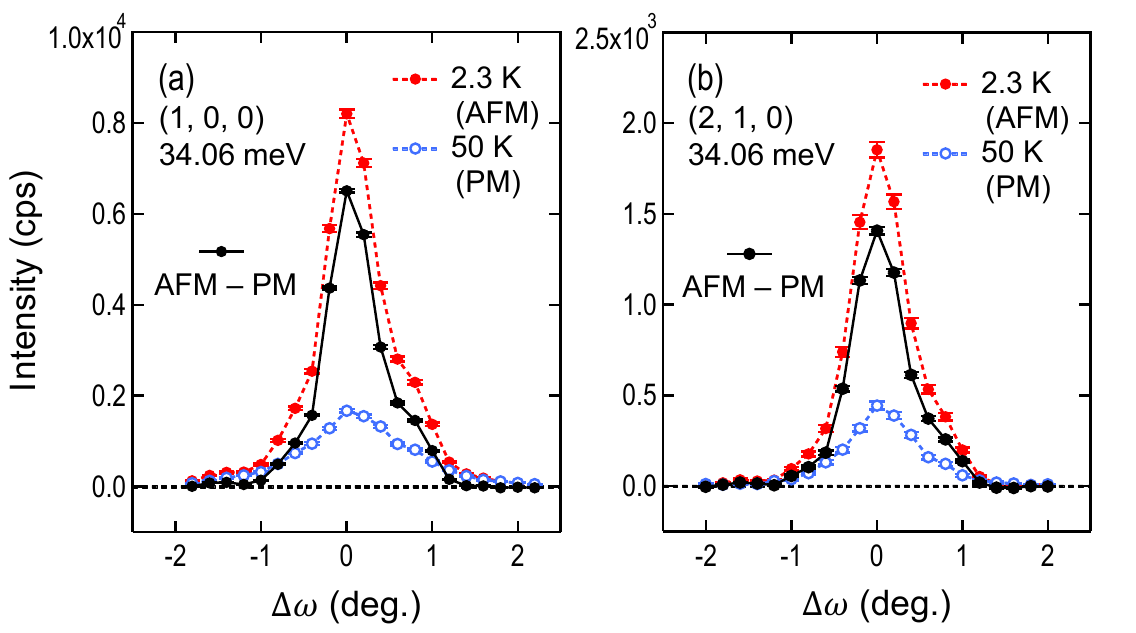}
\caption{(Color online) The peak profiles measured in the unpolarized mode for (a) (1, 0, 0) and (b) (2, 1, 0) reflections. The dashed lines with red symbols correspond to the data in the AFM phase at 2.3 K, and the dashed lines with blank blue symbols represent the data in the PM phase at 50 K. The solid lines with black symbols represent the net intensities obtained by subtracting the data at 50 K from those at 2.3 K.}
\label{Q0prof_unpol}
\end{centering}
\end{figure}

We started our experiments by confirming the AFM reflections with $\bm{Q} = 0$.
As reported in previous studies\cite{Steeman_1990, Ptasiewicz_1985, Sullow_2008}, we observed diffraction peaks in the AFM phase at the positions $\boldsymbol{\tau} =$ ($h$, $k$, 0) ($h + k = $ odd) that are forbidden for this crystal structure.
Figure \ref{Q0prof_unpol} shows typical peak profiles obtained from rocking curve measurements at the (1, 0, 0) and (2, 1, 0) reflection positions at 2.3 K and 50 K in the AFM and paramagnetic (PM) phases, respectively. 
Clear reflections were observed in the AFM phase at both positions, and their intensities significantly decreased upon heating to 50 K.
The residual signals at 50 K are attributed to nuclear reflections of 2$\boldsymbol{\tau}$, caused by the higher order contamination ($\lambda/$2) in the incident neutrons.
By subtracting this extrinsic contribution at 50 K as background, we analyzed the intrinsic magnetic scattering intensities that develop at low temperatures.
The net scattering profiles at 2.3 K are shown by black symbols with the solid lines in Fig. \ref{Q0prof_unpol}.
\begin{figure}[h]
\begin{centering}
\includegraphics[width=0.85\linewidth]{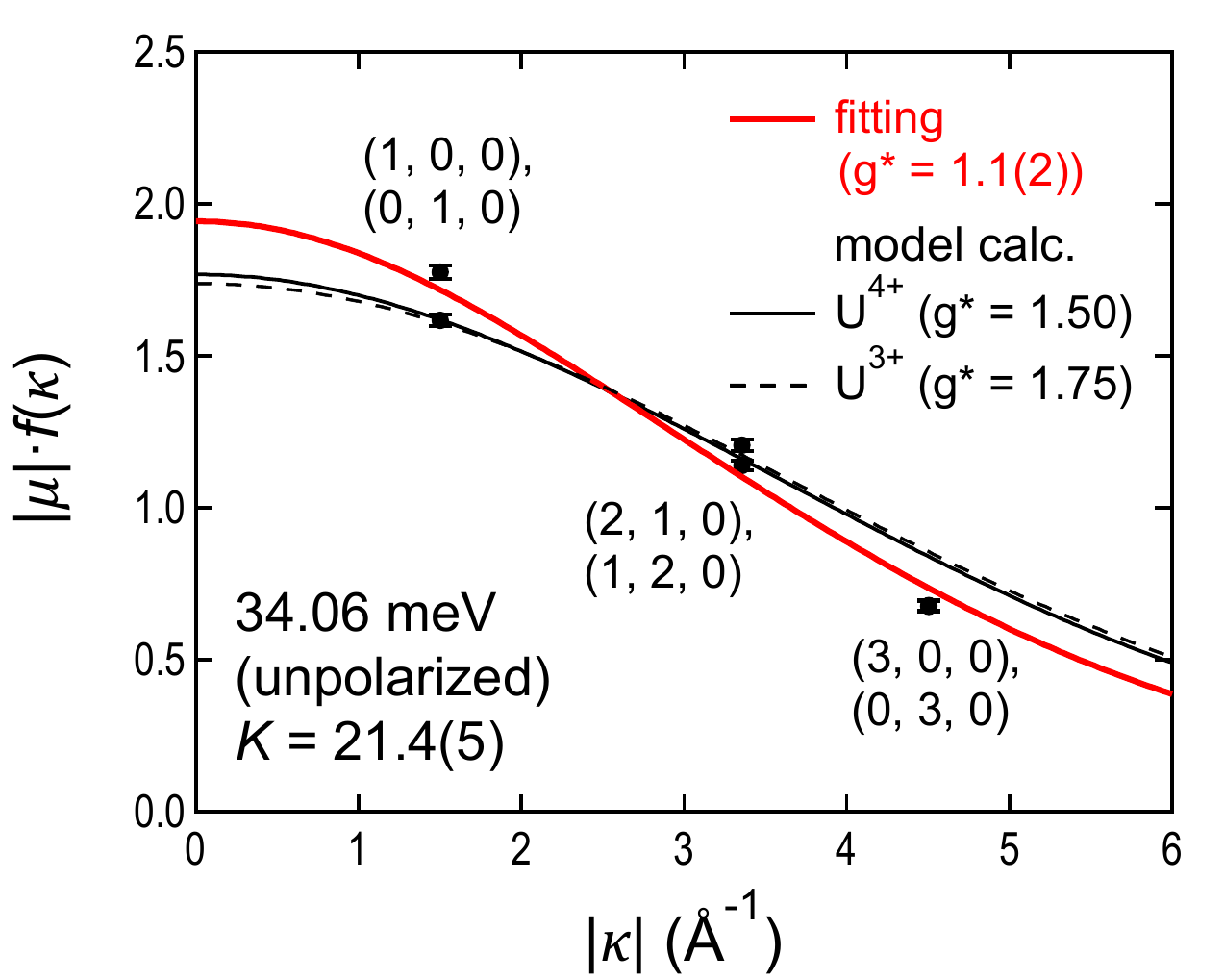}
\caption{(Color online) The values of $|\mu| \cdot f_{\mathrm{U}}$ estimated from the integrated intensities of magnetic reflections with $\bm{Q} = 0$, where $|\mu|$ represents the magnitude of magnetic moments, and $f_{\mathrm{U}}$ is the form factor of U.
The curved lines are the fitting results to the experimental data.
The thin solid and dashed black curves are based on assumptions for U$^{4+}$ and U$^{3+}$ ions, respectively.
The bold red cuve represents the fitting result using the parameters of $|\mu|$ and $g^*$ with the calculated functions of $\left\langle j_0(\kappa)\right\rangle$ and $\left\langle j_2(\kappa)\right\rangle$ for U$^{4+}$.}
\label{Q0_fmag}
\end{centering}
\end{figure}

The integrated intensity, $I_{mag}$, of a magnetic profile at the scattering vector $\boldsymbol\kappa$ is proportional to the square of ${\bm M}_{\perp} (\boldsymbol\kappa)$.
Here, ${\bm M}_{\perp} (\boldsymbol\kappa)$ refers to the projection of $\bm M(\boldsymbol\kappa)$ onto the plane perpendicular to $\boldsymbol\kappa$.
$\bm M(\boldsymbol\kappa)$ is the component of the Fourier-transformed magnetic-moment distributions with respect to $\boldsymbol\kappa$ (see Appendix B for more details).
$I_{mag}$ is expressed by the following equation: 
\begin{equation}
I_{mag}(\kappa) =K L(2\theta)\left|F_{m}(\kappa)\right|^2, 
\label{eq_I_F}
\end{equation}
where $K$ is the scale factor typically estimated from nuclear scattering intensities, 
$L(2\theta)$ is the Lorentz factor at the scattering angle of $2\theta$ for $\boldsymbol\kappa$, and $F_{m}(\kappa)$ is the magnetic structure factor per nuclear unit cell, defined as
\begin{equation}
\left|F_{m}(\boldsymbol{\kappa})\right|= b_{mag} \cdot\left|\boldsymbol{\mu}_{\perp}\right| \cdot f_{\mathrm{U}}(\kappa) \cdot A(\kappa).
\label{eq_F_mu}
\end{equation}
Here, $b_{mag}$ is the factor that converts the magnitude of the magnetic moment to the scattering length, a constant corresponding to the scattering length per 1$\mu_{\rm B}$ (2.965 fm).
$A(\kappa)$ represents the phase factor, and $|\boldsymbol{\mu}_{\perp}|$ is the magnitude of the projected vector of $\boldsymbol{\mu}$ onto the plane perpendicular to $\boldsymbol{\kappa}$.
For the AFM order with $\bm{Q} = 0$, where the ordered moments are parallel to the $c$-axis, these values are calculated as $A(\kappa) =$ 2 and $|\boldsymbol{\mu}_{\perp}| = |\boldsymbol{\mu}|$, respectively.
The magnetic form factor of U, $f_{\rm U}(\kappa)$, is a function that can be approximated by expanding it using an $l$'th order basis $j_{l}(\kappa)$ of spherical Bessel functions as follows:
\begin{align}
f_{\mathrm{U}}(\kappa)=\left\langle j_0(\kappa)\right\rangle+g^*\left\langle j_2(\kappa)\right\rangle,
\end{align}
where $g^*$ is defined as:
\begin{align}
g^*=\frac{J(J+1)+L(L+1)-S(S+1)}{3 J(J+1)-L(L+1)+S(S+1)}.
\end{align}
In this formula, $j_{0}(\kappa)$ is the term derived from the spin and orbital angular momentum of the electron, and $\braket{j_{2}(\kappa)}$ is the term derived purely from the orbital angular momentum.
These terms have been calculated using the Dirac-Fock method for the free ion model of U$^{4+}$ ($g^* =$ 1.50) and U$^{3+}$ ($g^* =$ 1.75)\cite{Desclaux_1978}.

These calculations are based on the assumption of isotropic electronic states.
However, in actual crystals, anisotropic electronic states may arise due to CEF effects.
In our analysis, following the approach commonly used for f-electron systems, the magnetic form factor was experimentally estimated using $g^*$ as a fitting parameter.
The magnetic scattering intensity observed at each reflection is converted into the product of $\mu$ and $f_{\rm U}(\kappa)$ using the Eqs. \eqref{eq_I_F} and \eqref{eq_F_mu}.
The resulting values are then plotted against $\kappa$, as shown in Fig. \ref{Q0_fmag}.
The thin solid and dashed black curves represent the fitting results using the calculated values of $f_{\rm U}(\kappa)$ for U$^{4+}$ and U$^{3+}$, respectively, with $|\boldsymbol{\mu}|$ being the only parameter.
There are no significant differences between the two models, and reproduce the experimental results quite well.
To further improve the agreement with the experimental data, $f_{\rm U}(\kappa)$ was experimentally determined by fitting both of $|\boldsymbol{\mu}|$ and $g^*$ as parameters.
The result is shown in Fig. \ref{Q0_fmag} with the red bold curve.
For this fitting, $j_{l}(\kappa)~(l = 0, 2)$ for U$^{4+}$ was used.
The obtained value of $\mu$ is 1.93(5) $\mu_{\rm B}$ ($g^* =$ 1.1(2)), which is consistent with the variation in values previously reported (1.67-2.5 $\mu_{\rm B}$/U)\cite{Steeman_1990, Ptasiewicz_1985, Sullow_2008}.

\begin{figure}[h]
\begin{centering}
\vspace{1\baselineskip}
\includegraphics[width=0.95\linewidth]{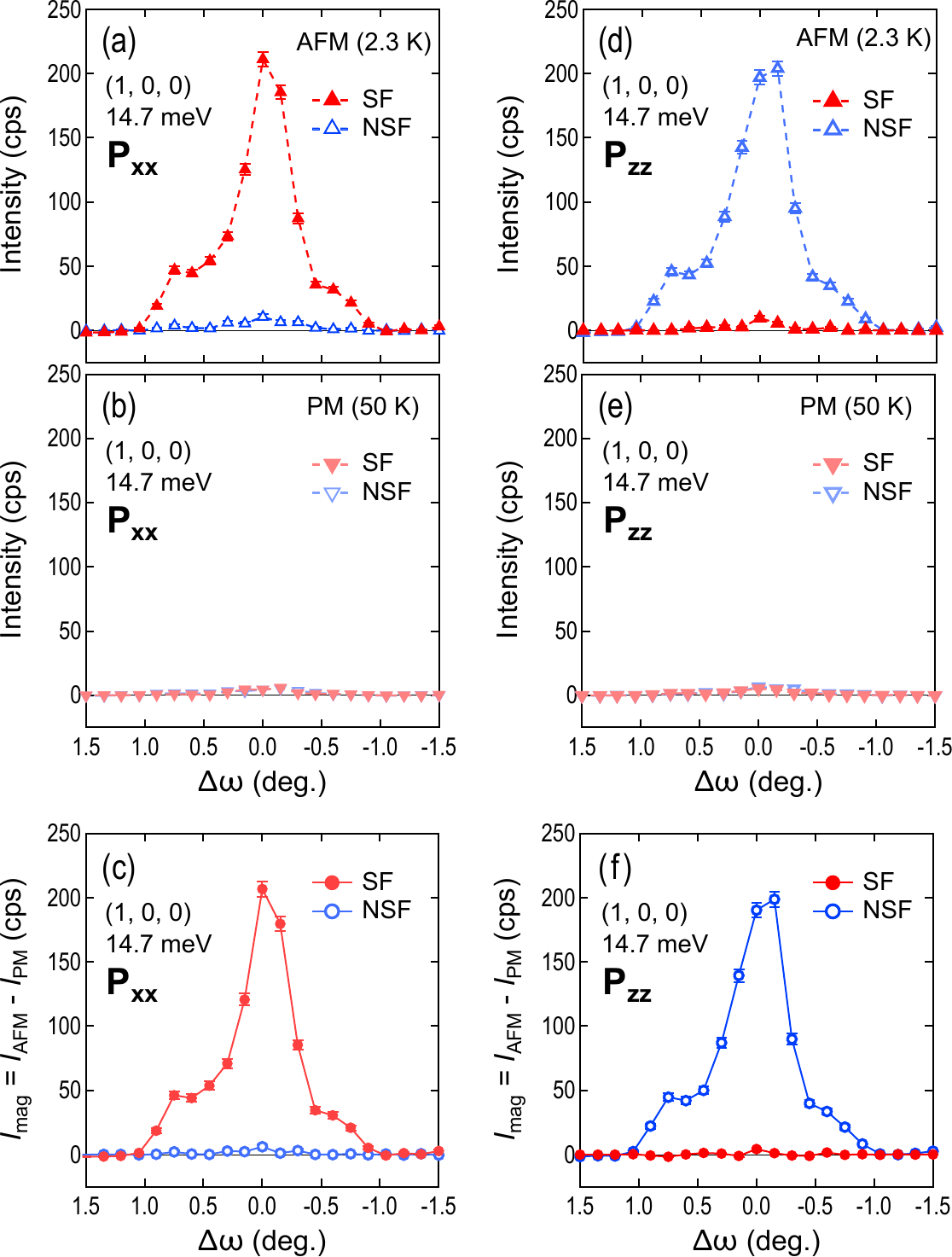}
\caption{(Color online) The peak profiles of SF and NSF scatterings for (1, 0, 0) reflection, measured in the polarized mode. (a) (d) Data of SF signal (represented by red{\large$\blacktriangle$} symbols) and NSF signal (blue{\large$\triangle$} symbols) measured at 2.3 K in the $P_{xx}$ and $P_{zz}$ settings. (b) (e) Data of SF signal (red{\large$\blacktriangledown$} symbols) and NSF signal (blue{\large$\triangledown$} symbols) measured at 50 K in the $P_{xx}$ and $P_{zz}$ settings. (c) (f) Data of SF signal (red{\large$\bullet$} symbols) and NSF signal (blue{\large$\circ$} symbols) obtained by subtracting the PM phase data from the AFM phase data in the $P_{xx}$ setting and $P_{zz}$ settings.}
\vspace{-1\baselineskip}
\label{Q0prof_pol}
\end{centering}
\end{figure}

To ascertain the direction of the moment, these signals were also investigated in polarized mode.
Figure \ref{Q0prof_pol} displays representative profiles from rocking curve measurements at the (1, 0, 0) reflection.
Specifically, Figs. \ref{Q0prof_pol}(a) and (b) present the peak profiles measured at 2.3 K (in the AFM phase) and 50 K (in the PM phase) in the $P_{xx}$ setting.
In this setting, nuclear and magnetic scatterings can be distinctly  observed as NSF and SF signals, respectively.
Regarding the SF signal, a prominent peak was observed in the AFM phase, which nearly vanished when the temperature reached the PM phase.
In contrast, no clear NSF signal was observed in either AFM and PM phases.
We extracted the magnetic scattering contribution by subtracting the PM phase data using the same approach as in the unpolarized mode.
The resulting magnetic signals are shown in Fig. \ref{Q0prof_pol}(c).
A distinct temperature-dependent variation is evident in the SF signal, while the NSF signal shows no such change.
This indicates that the observed variation in reflection intensity with temperature is solely attributed to magnetic scattering.

Rocking curve measurements for this reflection were also performed in the $P_{zz}$ setting to differentiate between the in-plane and $c$-axis components of the magnetic moment.
Figures \ref{Q0prof_pol}(d), (e) and (f) display their profiles in the AFM and PM phases, along with the extracted temperature differences, respectively.
Contrary to the results in the $P_{xx}$ setting, a clear temperature difference was observed only in the NSF mode.
The magnetic signal of SF scattering, corresponding to the $c$-plane component of the magnetic moment, was not detected within the experiment accuracy ($\lesssim$ 0.01 $\mu_{\rm B}$), which indicates that, for the $\bm{Q} = 0$ component, the magnetic moments are aligned parallel to the $c$-axis.
These observations are consistent with the results from previous neutron scattering experiments\cite{Steeman_1990, Ptasiewicz_1985, Sullow_2008} and our RXS experiments\cite{Kon_2023, Kon_2023_2}.
Similar measurements performed on other reflections, such as (0, 1, 0), (1, 2, 0), and (2, 1, 0), also yielded consistent results.

\subsection{$\bm{q}_{\rm CDW}$ reflections}
\begin{figure}[h]
\centering
\begin{minipage}[h]{0.9\linewidth}
\centering
\vspace{1\baselineskip}
\includegraphics[width=1\linewidth]{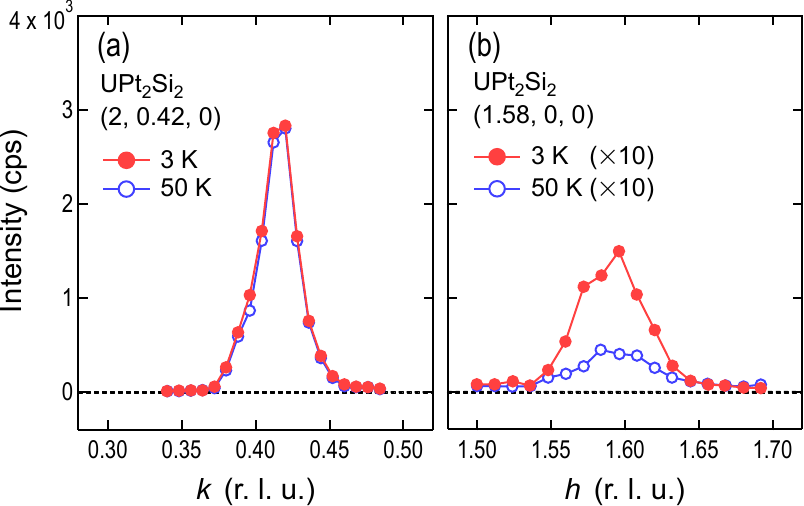}
\caption{(Color online) The diffraction peak profiles for (a) $k$-scan and (b) $h$-scan measured around the nuclear reflection of (2, 0, 0) in the unpolarized mode. The filled red symbols denote the data in the AFM phase at 3 K and the blank blue symbols represent the data in the PM phase at 50 K. Note that the vertical axis of graph (b) is scaled to 1/10 of that in (a).}
\label{qCDW_line}
\end{minipage}\\
\begin{minipage}[h]{1\linewidth}
\centering
\vspace{1\baselineskip}
\includegraphics[width=0.7\linewidth]{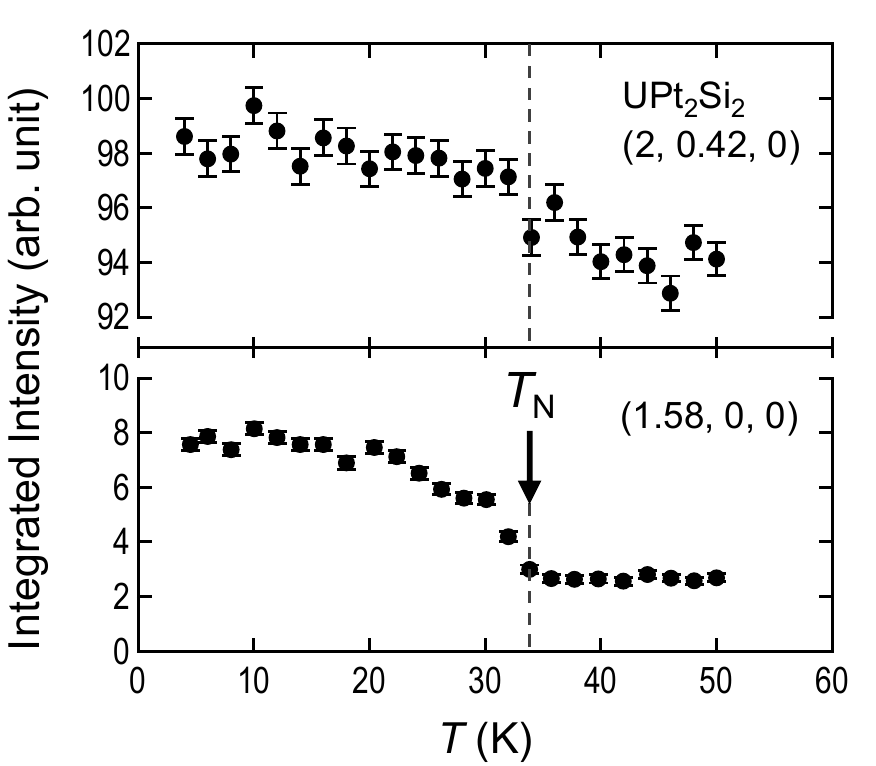}
\caption{The temperature dependence of the integrated intensities obtained in $\theta\mathchar`-2\theta$ scans for the (2, 0.42, 0) reflection (upper) and the (1.58, 0, 0) reflection (lower).}
\label{qCDW_Tdep}
\end{minipage}
\end{figure}

Next, we performed $h$, $k$ scans around the reciprocal lattice point (2, 0, 0) to search for superlattice reflections due to the CDW, in the unpolarized mode at 3 K. 
Figure \ref{qCDW_line} displays clear superlattice reflections observed at (2, $q_{\rm CDW}$, 0) and (2$-q_{\rm CDW}$, 0, 0).
There is a marked difference in the scattering intensities of these reflections, the latter being less than one-tenth the intensity of the former.
Upon raising the temperature to the PM phase, the already low scattering intensity of the (2$-q_{\rm CDW}$, 0, 0) reflection diminishes further, while the higher scattering intensity of (2, $q_{\rm CDW}$, 0) largely persists in the PM phase.

Figure \ref{qCDW_Tdep} presents the detailed temperature dependence of the integrated intensities for these reflections, obtained through $\theta\mathchar`-2\theta$ scans.
The scattering intensity of (2$-q_{\rm CDW}$, 0, 0) is observed to continuously increase below $T_{\rm N}$ with a distinct bend.
This increase in the intensity is considered to be due to a magnetic scattering resulting from the modulation of the AFM order as observed in our previous RXS measurements\cite{Kon_2023, Kon_2023_2}.

Although the data for the (2, $q_{\rm CDW}$, 0) reflection is somewhat noisy, a similar increase in intensity below around $T_{\rm N}$ is apparent.
This reflection position of (2, $q_{\rm CDW}$, 0) is, however, precisely where it would be affected by multiple scattering from the strong magnetic reflection at (1, 0, 0). 
Additionally, no signal was observed within the accuracy of experiments performed in the polarized mode using different incident wavelengths (refer to Appendix A).
We can thus conclude that the observed weak temperature variation in the intensity at (2, $q_{\rm CDW}$, 0) is not intrinsic.

The scattering intensities observed in the PM phase (blank blue circles in Fig. \ref{qCDW_line}(a) and (b)) are notable.
This is especially significant at (2, $q_{\rm CDW}$, 0) but very weak at (2$-q_{\rm CDW}$, 0, 0). 
In fact, we have ascertained that the latter is an extrinsic signal resulting from multiple scatterings of the nuclear reflection at (3, 1, 0) (refer to Appendix A). 
Thus, it can be stated that around (2, 0, 0), CDW satellite reflections are observed only in the $k$ direction. 
This result aligns with Lee $et~al$.'s observations \cite{Lee_2020}, where satellites are detected in the $k$ direction for ($h$, 0, 0) ($h \neq$ 0) and in both $h$ and $k$ directions for ($h$, $k$, 0) ($h \neq$ 0, $k \neq$ 0).

\begin{figure}[h]
\begin{centering}
\includegraphics[width=0.8\linewidth]{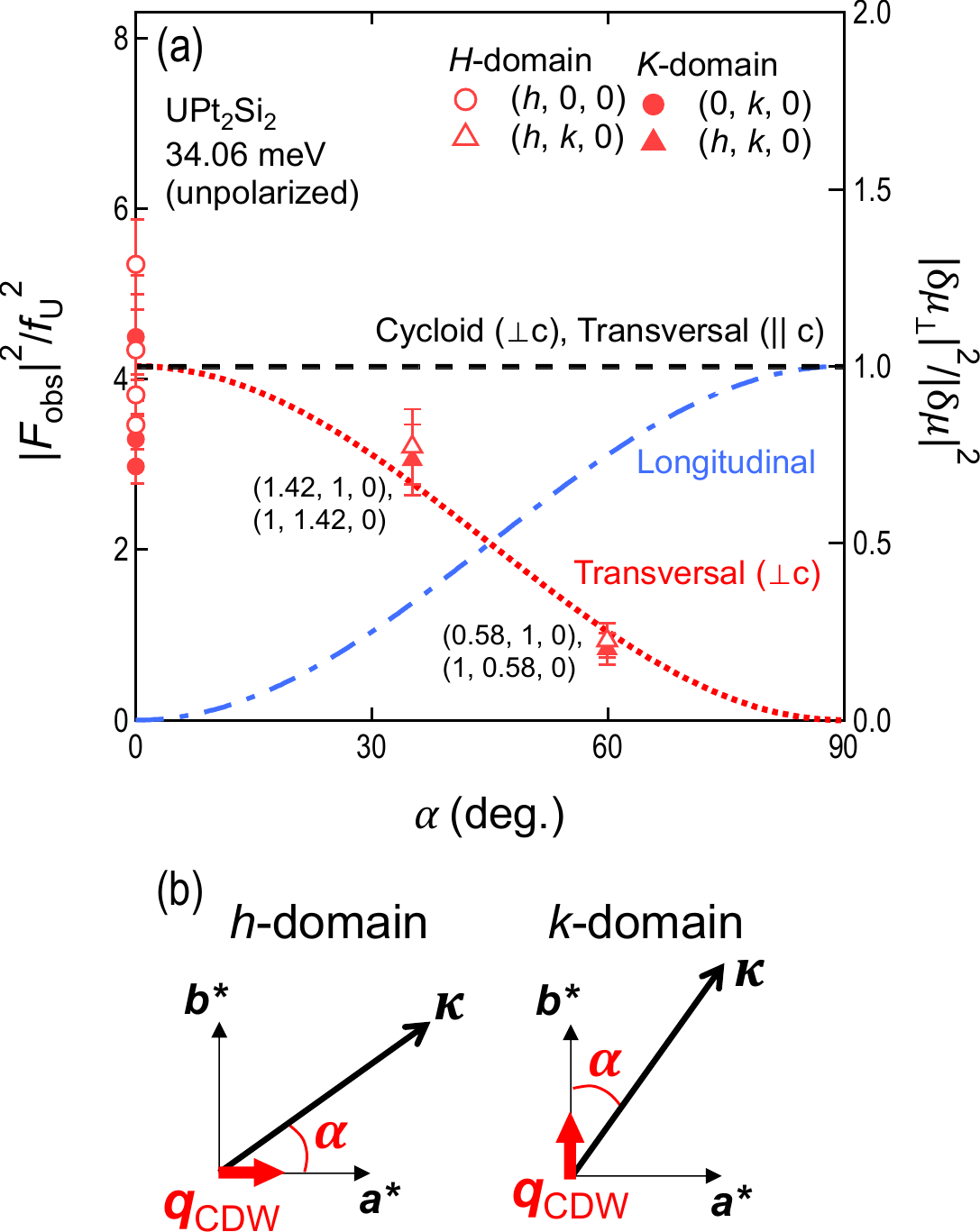}
\caption{(Color online) (a) On the left axis, the values of $\left|F_{\rm obs}\right|^2/f_{\rm U}(\kappa)^2$ derived from the observed magnetic scattering intensities are plotted with the symbols. The filled symbols and the blank ones correspond to the reflections of $h$-domain and $k$-domain, respectively. Plotted on the right axis are the calculated curves of $|\boldsymbol{\delta\mu}_{\perp}|^2/|\boldsymbol{\delta\mu}|^2$ for different modulations: the $c$-plane transverse modulation (red dotted line), the longitudinal modulation (blue dashed-dotted line), and the cycloid modulation within the $c$-plane or the transverse modulation parallel to the $c$-axis (black dashed line). (b) The definition of $\alpha$ for the superlattice reflections of $h$-domain and $k$-domain.}
\label{qCDW_alpha}
\end{centering}
\end{figure}

To identify the magnetic modulation structure, we performed rocking curve measurements on the superlattice reflections within the accessible range in the ($hk$0) plane and made a quantitative analysis of the magnetic contributions to their integrated intensities.
Each of the magnetic scattering intensities observed can be transformed into a magnetic structure factor per nuclear unit cell, $\left|F_{\rm obs}\right|$, using Eq. \eqref{eq_I_F}, similar to the $\bm{Q} = 0$ case.
To find the direction of the magnetic moments $\boldsymbol{\delta\mu}$, which composes the $\bm{q}_{\rm CDW}$ magnetic modulation, we first evaluated the variation of $\left|F_{\rm obs}\right|^2/f_{\rm U}(\kappa)^2$ with respect to the scattering vector $\boldsymbol{\kappa}$.
This quantity is proportional to the square of $|\boldsymbol{\delta\mu}_{\perp}|/|\boldsymbol{\delta\mu}|$, where $\boldsymbol{\delta\mu}_{\perp}$ represents the projection of the $\boldsymbol{\delta\mu}$ onto the plane perpendicular to $\boldsymbol{\kappa}$.
Figure \ref{qCDW_alpha}(a) shows $\left|F_{\rm obs}\right|^2/f_{\rm U}(\kappa)^2$ on the left axis and the calculated values of $|\boldsymbol{\delta\mu}_{\perp}|^2/|\boldsymbol{\delta\mu}|^2$ for various magnetic modulation models on the right axis, plotted against the angle $\alpha$.
This angle, $\alpha$, is defined as the angle between $\bm{q}_{\rm CDW}$ and $\boldsymbol{\kappa}$ at each reflective position, as illustrate in Fig. \ref{qCDW_alpha}(b).
For example, in transverse modulation structures within the $c$-plane, $|\boldsymbol{\delta\mu}_{\perp}|$ is proportional to $\cos\alpha$,
while it is proportional to $\sin\alpha$ in longitudinal magnetic modulations.
For a transverse wave where only the $c$-axis magnetic moments modulate, or a cycloidal structure within the $c$-plane, $|\boldsymbol{\delta\mu}_{\perp}|$ is constant.
More intricate structures, like conical modulations, can be understood as combinations of these three scenarios.
Our observations show that $|\boldsymbol{\delta\mu}_{\perp}|$ significantly decreases in a manner consistent with $\cos\alpha$ as $\alpha$ increases, aligning well with the transverse magnetic modulation model in the $c$-plane. 
Thus, we conclude that the magnetic structure of this system incorporates a transverse magnetic modulation component in the $c$-plane, with a periodicity of ${q}_{\rm CDW}$, leading to a deviation of the $\bm{Q} =$ 0 ordered moments from the $c$-axis. 
This conclusion is consistent with the structure we proposed from the RXS experiments.

\begin{figure}[h]
\vspace{1\baselineskip}
\begin{centering}
\includegraphics[width=0.85\linewidth]{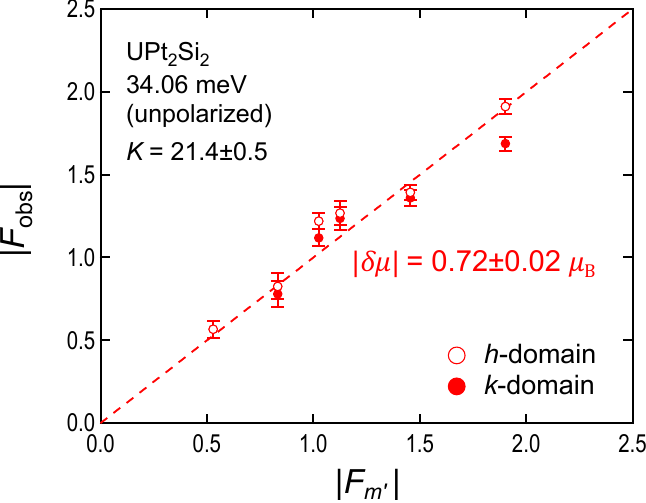}
\caption{(Color online) Comparison of the observed magnetic structure factor $F_{\rm obs}$, derived from magnetic scattering intensities, with the calculated $F_{m'}$ assuming the transverse magnetic modulation. Filled symbols and blank symbols represent the reflections of $h$-domain and $k$-domain, respectively.
The dashed line denotes $\left|F_{\rm obs}\right|/\left|{F_{m'}}\right| = 1$.}
\label{qCDW_FF}
\end{centering}
\end{figure}

Figure \ref{qCDW_FF} displays a comparison between $\left|F_{\rm obs}\right|$ and $\left|F_{m'}\right|$, where $\left|F_{m'}\right|$ is the magnetic structure factor calculated for the $c$-plane transverse magnetic modulation:
\begin{equation}
\left|F_{m'}(\boldsymbol{\kappa})\right|= b_{mag} \cdot\left|\boldsymbol{\delta\mu}_{\perp}\right| \cdot f_{\mathrm{U}}(\kappa)
\label{Fmag_CDW}
\end{equation}
(refer to Appendix B).
A good agreement between $\left|F_{\rm obs}\right|$ and  $\left|F_{m'}\right|$ is achieved when $\left|F_{m'}\right|$ is calculated using $|\boldsymbol{\delta\mu}| =$ 0.72(2) $\mu_{\rm B}$/U.
Notably, this value exceeds one-third of the amplitude for the $\bm{Q} = 0$ component, 1.93(5) $\mu_{\rm B}$/U, determined in this study.
This means that the ordered magnetic moments are tilted from the $c$-axis at a maximum angle of $\sim$ 20$^\circ$.


\begin{figure*}[h]
\begin{centering}
\vspace{1\baselineskip}
\includegraphics[width=0.9\linewidth]{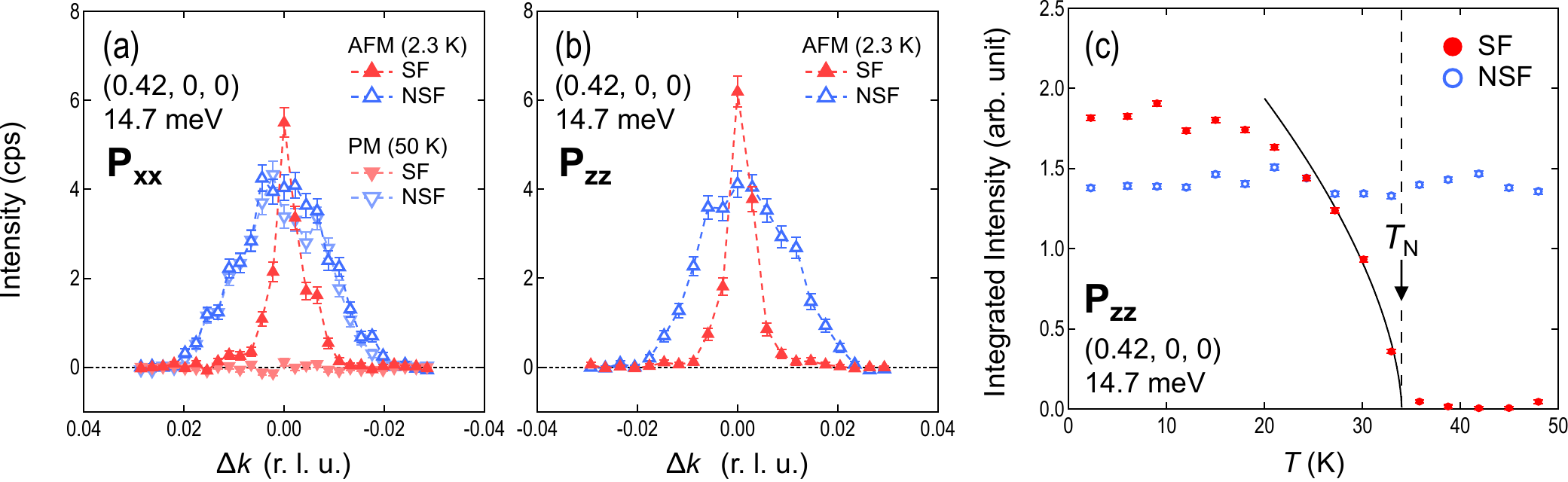}
\caption{(Color online) The peak profiles and the temperature dependence of SF (filled symbols) and NSF (open symbols) signals of the (0.42, 0, 0) reflection measured in the polarized mode.
(a) The profiles measured in the $P_{xx}$ setting at 2.3 K ({\large$\blacktriangle$} and {\large$\triangle$}) and at 50 K ({\large$\blacktriangledown$} and {\large$\triangledown$}). (b) The profiles measured in the $P_{zz}$ setting at 2.3 K ({\large$\blacktriangle$} and {\large$\triangle$}). (c) The temperature dependence of the integrated intensity measured in the $P_{zz}$ setting. The filled circles denote the data for SF signals and the blank blue circles represent the data for NSF signals. The black curved line represents the fitting result for the SF scattering intensity near the $T_{\rm N}$, using the equation of $I_{mag} \propto |T - T_{\rm N}|^{2\beta}$ ($\beta=0.3$).}
\label{qCDW_prof_pol}
\end{centering}
\end{figure*}

To further confirm the determined magnetic modulation structure, rocking curve measurements in the polarized mode were performed for several reflections.
Figure \ref{qCDW_prof_pol} features representative profiles of the (0.42, 0, 0) reflection.
In the $P_{xx}$ setting, a broad NSF signal without intensity change across $T_{\rm N}$ and a sharp SF signal observable only in the AFM phase are noted (Fig. \ref{qCDW_prof_pol}(a)).
As per Table \ref{polmode}, the former is attributed to nuclear scattering, and the latter to magnetic scattering in this setting.
To distinguish between the magnetic modulation contributions along the $c$-axis and in the $c$-plane in the observed magnetic scattering, the $P_{zz}$ setting was employed.
Here, NSF signals correspond to $c$-axis modulation, while SF signals indicate $c$-plane modulation.
As Fig. \ref{qCDW_prof_pol}(b) shows, both NSF and SF signals exhibit no difference from those in the $P_{xx}$ setting.
The temperature dependence of integrated intensities of these signals is detailed in Fig. \ref{qCDW_prof_pol}(c).
The SF signal is present only below $T_{\rm N}$, while the NSF signal shows no notable temperature variation around $T_{\rm N}$.
There was no significant temperature change in the peak width of the magnetic signals.
These observations confirm that the magnetic modulation predominantly involves the $c$-plane component of the magnetic moments, with negligible $c$-axis component contribution within the experimental accuracy ($\lesssim$ 0.01 $\mu_{\rm B}$).
Similar results are obtained for other $\bm{q}_{\rm CDW}$ superlattice reflections like (0.58, 1, 0) and (1.58, 0, 0).

\section{Discussion}
Now, we discuss how the CDW induces magnetic modulation in \ce{UPt2Si2}. 
Similar phenomena have been observed in the 4f electron system of \ce{TbTe3}\cite{Lee_2012, Chillal_2020}. 
Like \ce{UPt2Si2}, \ce{TbTe3} exhibits a quasi-two-dimensional layered structure, with the 5p electrons of Te forming a CDW order below $\sim$ 330 K and the 4f electrons of Tb undergoing an AFM order below $\sim$ 6.6 K. 
The AFM structure is modulated with the same period as the CDW, causing the principal axis of the magnetic moment to tilt by up to $\sim$ 5$^\circ$. 
This magnetic modulation is considered to be attributed to the spatial modulation of the CEF levels, impacting the well-localized 4f electrons in Tb.
Specifically, the amplitude of modulation in the CEF level splitting is estimated to be $\sim$ 5 meV, from the RXS measurements \cite{Lee_2012}.
This CEF modulation, involving changes in the shape of the 4f electron orbitals, results in the modulation of the direction of the magnetic moments.
Namely, the CDW induces an orbital order of the 4f electrons, which, through spin-orbit interaction, also modulates the magnetic order. 
The essential element of this scenario is the CEF modulation, which is presumed to arise from the displacement of Te ions and an approximate 0.3\% change in the Te-Tb interionic distance associated with the CDW order.

In the case of \ce{UPt2Si2}, a similar scenario might be applicable when viewed from a
localized perspective. 
According to the studies by Lee $et~al.$\cite{Lee_2020}, although based on a two-dimensional analysis of limited scattering plane data, the primary displacement in Layer 2 is of Pt(2) atoms, induced by the CDW. 
The direction of this displacement is mainly in the direction perpendicular to the CDW propagation direction ([100] or [010]), with an amplitude estimated at $\sim$ 0.15 \AA. 
This results in a maximum change of about 3.0\% in the U-Pt(2) bond length. 
However, assuming the valence of Pt ions remains unchanged, the displacement of Pt causes a disarray of at most 2$^\circ$ in the principal axis direction at the U site, which seems insufficient to explain the observed 20$^\circ$ tilt of the magnetic moments. 
Therefore, to explain the experimental results based on a CEF model assuming effective point charges, a significant reconfiguration of charge among Pt ions adjacent to the $\bm{q}_{\rm CDW}$ in a perpendicular direction is anticipated. 
However, no such additional CDW induction in this direction has been observed. 
By contrast, in \ce{TbTe3}, although no quantitative discussion has been made regarding the relationship between the CEF modulation and the tilt angle, the emergence of CDW coexisting and orthogonal to the main CDW wave vector at low temperatures is noteworthy.

\begin{figure}[h]
\begin{centering}
\vspace{1\baselineskip}
\includegraphics[width=0.9\linewidth]{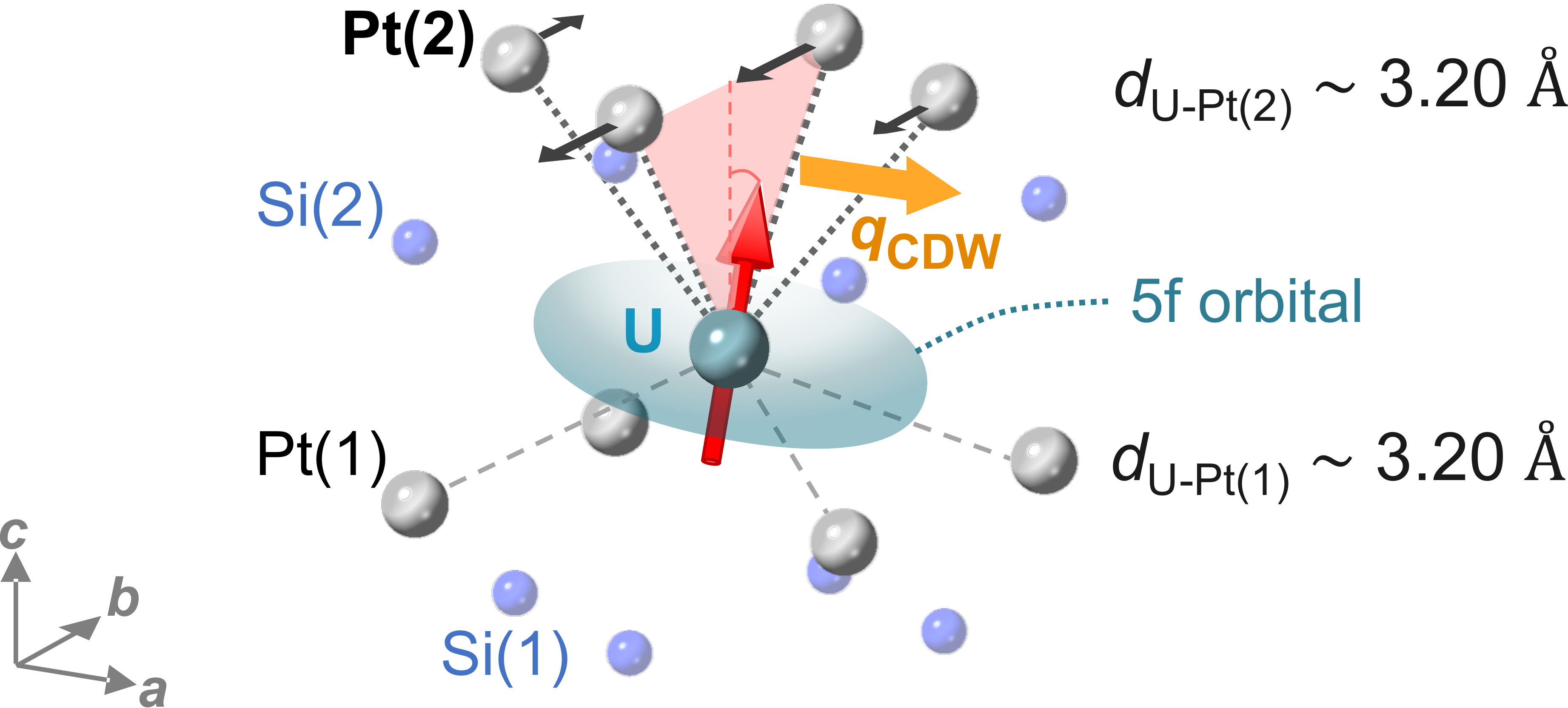}
\caption{(Color online) An illustration depicting the relationship between a single uranium ion and surrounding ions in \ce{UPt2Si2}. The black arrows orthogonal to $\bm{q}_{\rm CDW}$ represent the primary direction of displacement of Pt(2) ions induced by the CDW order. The red arrows illustrate the tilting of the uranium magnetic ordered moment in a plane orthogonal to $\bm{q}_{\rm CDW}$, as revealed in this experiment. The shaded area represents a conceptual illustration of the 5f orbitals contributing to the d-f antibonding orbital.}
\label{Env_U}
\end{centering}
\end{figure}

On the other hand, it is significant  that the observed tilt angle of the magnetic moment is roughly aligned with the angle $\sim$ 40$^\circ$ formed by the bond between Pt(2) and U with the crystallographic $c$-axis. 
As depicted in Fig.~\ref{Env_U}, the U ions in \ce{UPt2Si2} are in close proximity to four Pt(2) ions located within the same $ac$ and $bc$ planes, and to four Pt(1) ions positioned at a 45-degree offset. 
The bond lengths for both are approximately 3.2 \AA. 
Our observations indicate that the magnetic moment tilts from the $c$-axis towards the $b$-axis in response to the $\bm{q}_{\rm CDW}$ aligned along the $a$-axis. 
In the CDW state, the displacement of one of the two Pt(2) ions, aligned in the $b$-axis direction, towards U \cite{Lee_2020} suggests a selective bonding of the U 5f orbital with this particular Pt. 
This selective bonding, while competing with the primary AFM correlations, can naturally
explain the substantial tilt of the principal axis. 
The experimental results imply that the d-f hybridization between U and the four Pt(2) ions is highly sensitive to the breaking of fourfold symmetry caused by the CDW, significantly disrupting its balance.

The above discussion emphasizes the role of hybridization not just as a perturbation but as a crucial component of molecular orbitals, based on the nature of 5f electrons having spatially more extended wavefunctions than 4f electrons. 
The d-f hybridization lowers the energy of 5d orbitals to form bonding orbitals, while raising the energy of 5f orbitals to form anti-bonding orbitals.
As a result, the 5f orbital with the lowest energy having less overlap with the Pt(2) 5d orbitals, tend to spread orthogonally to the U-Pt(2) bond, thus orienting the magnetic moments towards the bond's direction, as conceptually illustrated in Fig.~\ref{Env_U}. 
The actual tilt angle seems to be determined by the competition between this effect and the main AFM correlations. 
Such behavior in solids, akin to molecular orbitals, is often observed in d-electron systems
with more extensive wavefunctions, resembling what is known as nematic order. 
For example, in \ce{IrTe2} \cite{Takubo_2014}, the CDW induced by Te 5p electrons modulates the 5d orbitals of Ir through the hybridized bonding orbitals with Te 5p, resulting in the modulation of 5d orbitals (stripe order), similar to the behavior observed in \ce{UPt2Si2}.

The atomic displacements in this system have also been discussed in recent pair-distribution-function (PDF) analyses by Petkov $et~al$\cite{Petkov_2023}. 
According to their study, consistent with the reports by Lee $et~al$.\cite{Lee_2020}, the primary atomic displacements are attributed to Pt(2), with the magnitude of displacement being $\sim$ 0.15 \AA. 
Their PDF analyses further revealed two notable new findings: First, not only Pt(2) but also U undergoes a small in-plane displacement of about 0.03 \AA. 
The direction of displacement is not [100] but [110], and although small compared to the in-plane displacement, there is also displacement perpendicular to the plane. 
Second, and more notably, the displacement in which the sheets of Pt(2) and U alternately shift in the [110] direction occurs uniformly at higher temperatures than the $T_{\rm CDW}$. 
No displacements of other atoms have been detected, suggesting that the bond between U and Pt within this crystal structure partially breaks the fourfold symmetry, leading to a structural instability. 
While the details of atomic displacement require further elucidation through precise structural analysis, their analysis supports our inference that the selective bonding between U and Pt(2)
is induced by the local symmetry breaking.

We also note that, in Fig.~\ref{qCDW_prof_pol}, the polarized neutron scattering peak profiles in the $k$-direction at $\bm{q}_{\rm CDW}$ reveal a distinct difference in the transverse peak widths between the SF and NSF scattering modes. 
The resolution, inferred from the $\kappa$-dependence of nuclear scattering peak widths at fundamental reflection positions, aligns with that of SF scattering, indicating that SF scattering nearly reaches the resolution limit. 
Conversely, NSF scattering peaks are markedly broader, with an estimated correlation length of $\sim$ 70 \AA~(about 17$a$). 
Hence, it was concluded that while the CDW order in the transverse direction is of short-range, the incommensurate magnetic order induced in the U atom layers is long-range. 
This is likely due to the strong ferromagnetic correlations within the U atom layers of the main $\bm{Q} =$ 0 AFM order, which remain undisturbed in the direction perpendicular to the propagation of magnetic modulation. 
In $h$-direction scans, on the other hand, no significant peak width differences were noted (not shown). 
Since the resolution in this direction is undeterminable from the present experimental data, discussing the correlation lengths for both orders is not feasible, yet they are likely to be similar in magnitude.

Moreover, it raises a compelling question: does the selective bonding between U and
Pt(2) enhance or conflict with the RKKY interaction derived from d-f hybridization? 
In a localized model, magnetic interactions that establish $\bm{Q} = 0$ order are typically linked to RKKY interactions. 
However, the specific orbitals of conduction electrons mediating this process remains elusive, even in first-principles calculations\cite{Elgazzar_2012}. 
Conversely, the presence of CDW order points to nesting instabilities in the Fermi surface of 5d
electrons, thereby fostering favorable conditions for the formation of a SDW at the same $\bm{q}_{\rm CDW}$ wave number. 
This phenomenon is exemplified in the 4f electron system of \ce{GdNiC2} \cite{Hanasaki_2017}, where a phase shift in spin density under CDW order gives rise to SDW, possibly correlating with the 4f electron magnetism. 
Future research, particularly employing RXS of Pt 5d electrons (including under magnetic fields), to explore the existence of SDW associated with Pt 5d and their magnetic interplay with 5f  electrons, will offer intriguing insights.

Finally, from the perspective of the itinerant model, the hybridization between U 5f and
Pt 5d is interpreted as band hybridization. 
According to first-principles calculations\cite{Elgazzar_2012}, the 5f orbitals exhibit itinerant characteristics near the Fermi energy, while the 5d electrons of Pt form a valence band centered around $-$4.5eV, but also hybridize with the 5f electrons, contributing to the density of states at the Fermi level. 
The Fermi surface formed by these hybridized bands reflects the two-dimensional nature of the layered structure. 
These theoretical calculations were conducted before the discovery of CDW order and do not account for the nesting instabilities of the CDW wave number nor the spontaneous breaking of symmetry in the 5f-5d orbital hybridization. 
Future studies exploring whether more stable electronic and crystal structures exist near this crystal structure will be a fascinating direction of research.

\section{Conclusion}
In this study, we carried out polarized and unpolarized neutron diffraction experiments on \ce{UPt2Si2} to thoroughly investigate the magnetic structure of its AFM phase. 
We reconfirmed the previously reported $\bm{Q} = 0$ AFM order with the ordered magnetic moment estimated to be 1.93(5) $\mu_{\rm B}$/U, aligned parallel to the $c$-axis within an experimental accuracy of $\sim$ 0.01 $\mu_{\rm B}$/U. 
Moreover, we observed nuclear scattering corresponding to atomic displacements induced by the CDW order and magnetic modulation waves occurring at the same periodicity as the CDW, $\bm{q}_{\rm CDW} =$ ($\sim$0.42, 0, 0). 
These findings are in line with our previous RXS experiments. 
A significant new insight gained is the amplitude of the magnetic modulation, determined to be 0.72(2) $\mu_{\rm B}$/U, suggesting a tilting of U magnetic moments by about 20$^\circ$ due to the CDW influence. The CDW-induced average structural changes in the Pt(2) atoms are known to be small, indicating that the tilting of 5f electron orbitals is not adequately accounted for by straightforward CEF modulation alone.
Based on these results, we propose that selective hybridization between U 5f electrons and adjacent Pt(2) 5d electrons spontaneously breaks the fourfold symmetry of 5f orbitals, significantly tilting their principal axis. 
The quantitative data obtained on the interactions between U 5f and Pt 5d electrons in \ce{UPt2Si2} is expected to contribute to the future progress in theoretical studies, including first-principles calculations, for a microscopic
understanding of the role of d-f hybridization in the 1-2-2 series. 
Experimentally, detailed crystallographic analysis and direct observation of Pt 5d electron orbitals through RXS,
especially including the high-temperature range above $T_{\rm CDW}$, are anticipated to provide further insights.

\begin{acknowledgment}
\acknowledgment
{\small \bf{Acknowledgement}} {\footnotesize The authors deeply appreciate H. Nakao, H. Kusunose, S. Hayami, H. Harima, and S. S\"{u}llow for valuable discussions. This work was supported by JSPS KAKENHI Grant Numbers JP15H05882, JP15H05883
(J-Physics), JP23H04867, and JP23H04866 (Asymmetry Quantum Matters).
The neutron scattering experiments were carried out by the joint research in the Institute for Solid State Physics, the University of Tokyo (No. 22528).
}
\end{acknowledgment}

\appendix
\vspace{2\baselineskip}
\section{The diffraction pattern of superlattice reflections with $\bm{q}_{\rm CDW}$ by nuclear and magnetic scatterings}
\begin{figure}[h]
\centering
\includegraphics[width=1\linewidth]{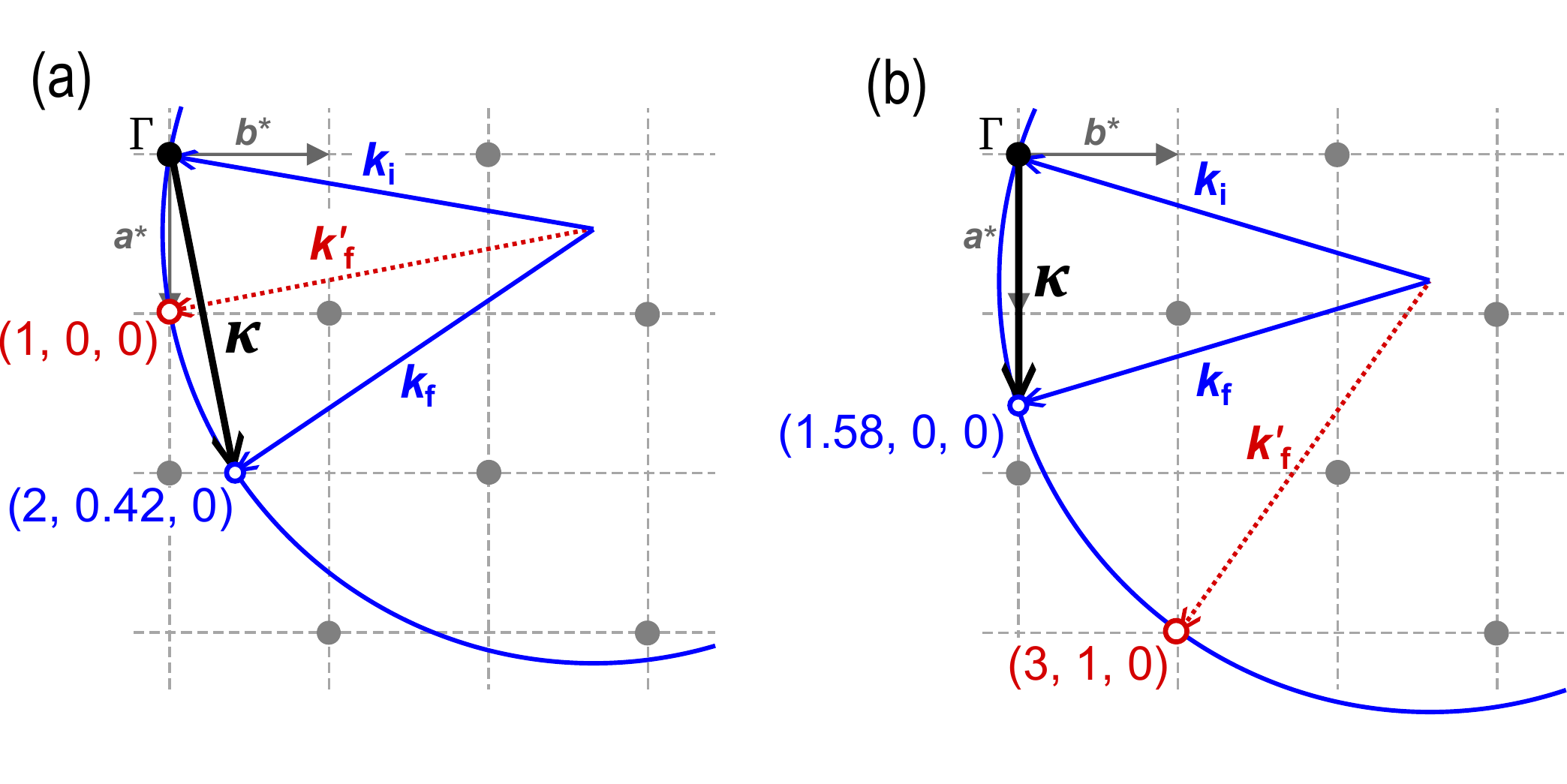}
\caption{(Color online) The Ewald spheres on the ($hk$0) plane associated with the reflections at (a) (2, 0.42, 0, 0) and (b) (1.58, 0, 0) in the unpolarized mode ($\lambda =$ 1.55 $\AA$, $k =$ 4.05 $\AA^{-1}$) are represented by blue solid-line circles. The blue thin arrows indicate the wave vectors of the incident ($\bm{k}_{\rm i}$) and scattered ($\bm{k}_{\rm f}$) neutrons in the Bragg conditions for the target reflections. The scattering vectors are denoted by the black bold arrows.
The red circle in each panel denotes a reflection present on the Ewald sphere besides the target reflection.
Scattering from $\bm{k}_{\rm i}$ to both $\bm{k}_{\rm f}$ and $\bm{k}'_{\rm f}$ occurs simultaneously. Additionally, $\bm{k}'_{\rm f}$ can cause multiple scattering events in the sample, such as diffracting to $\bm{k}_{\rm f}$ or diffracting back to $\bm{k}_{\rm i}$ and then to $\bm{k}_{\rm f}$. These processes can lead to a non-essential enhancement of the scattering intensity in the target reflections. 
}
\label{RS_multi}
\vspace{-1\baselineskip}
\end{figure}
\begin{figure}[h]
\centering
\includegraphics[width=0.8\linewidth]{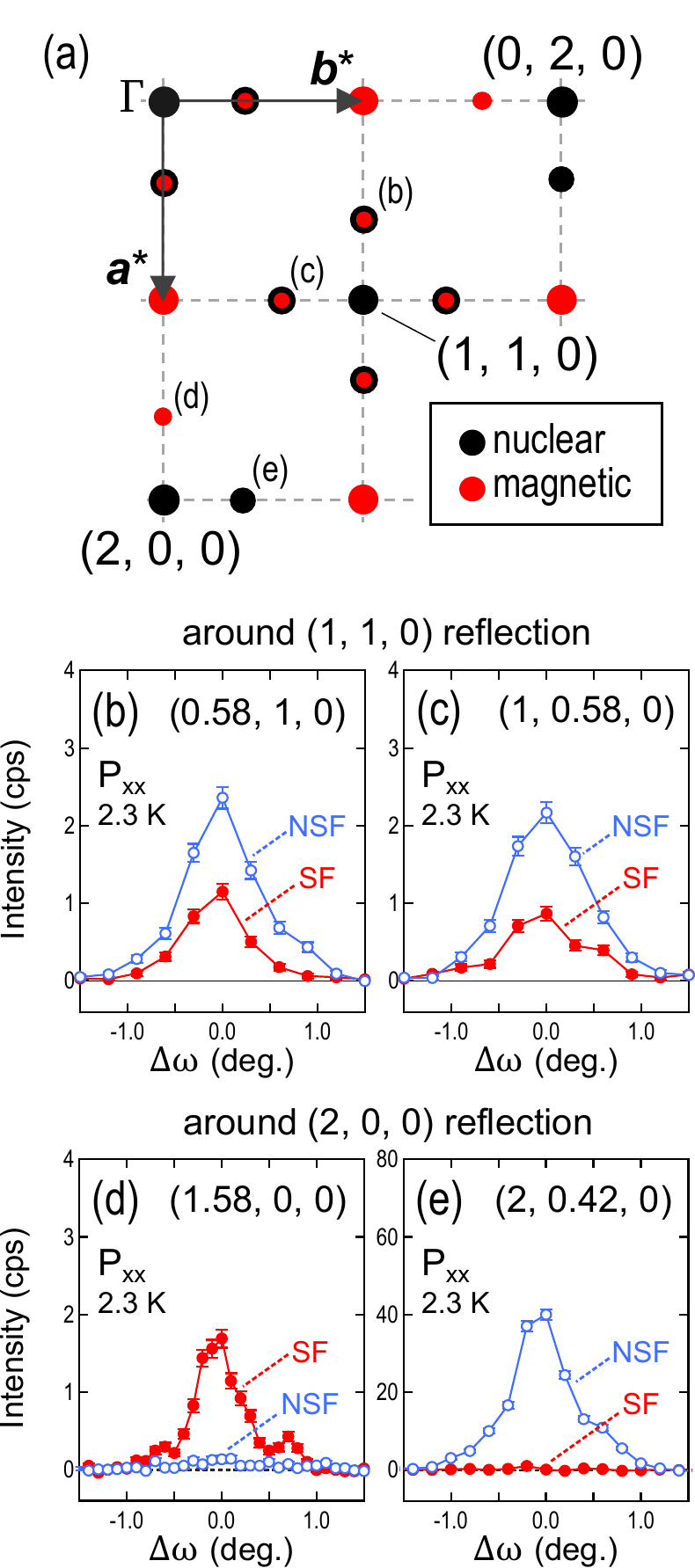}
\caption{(Color online) (a) The map of the diffraction pattern in the ($hk$0) plane, drawn based on the results in the polarized mode. The black and red circles indicate the nuclear and magnetic reflections, respectively. (b)-(e) The peak profiles of SF and NSF signals at the superlattice reflections characterized by $\bm{q}_{\rm CDW}$, measured in the polarized mode with the $P_{xx}$ setting. The profiles of the reflections around (1, 1, 0) and (2, 0, 0) nuclear Bragg peaks are shown in panels (b), (c) and (d), (e), respectively.}
\label{RS_prof}
\vspace{-1\baselineskip}
\end{figure}
Figure \ref{RS_multi} shows the scattering conditions in the unpolarized mode for the satellite reflections at (2, 0.42, 0) and (1.58, 0, 0).
On the Ewald sphere corresponding to the (2, 0.42, 0) reflection, the (1, 0, 0) magnetic Bragg point is also located (Fig. \ref{RS_multi}(a)).
Likewise, for the (1.58, 0, 0) reflection, it coincides with the (3, 1, 0) nuclear Bragg point (Fig. \ref{RS_multi}(b)).
In these circumstances, Bragg scattering simultaneously occurs from the incident X-ray beam $\bm{k}_{\rm i}$ to both $\bm{k}_{\rm f}$ and $\bm{k}'_{\rm f}$. 
Furthermore, a new diffracted wave, considered as originating from $\bm{k}'_{\rm f}$ as the incident wave, along with its successive repetitions, superimposes on $\bm{k}_{\rm f}$, leading to a so-called multiple-scattering effect. 
Consequently, the observed increase in intensity below $T_{\rm N}$ at (2, 0.42, 0) and the signal in the PM phase detected at (1.58, 0, 0) can be ascribed to multiple scattering from the (1, 0, 0) magnetic reflection and the (3, 1, 0) nuclear reflection, respectively.

In the main text, we explained that the unpolarized mode, using neutrons with shorter wavelengths, is more prone to multiple reflections and scatterings compared to the polarized mode. 
Therefore, our discussion on the diffraction patterns of both nuclear and magnetic scatterings in this system, primarily relies on the polarized mode results.
The obtained patterns are shown in Fig.~\ref{RS_prof}(a), 
with the panels (b)-(e) displaying the profiles of the superlattice reflections identified in Fig.~\ref{RS_prof}(a).
All of these profiles, measured in $P_{xx}$ setting, enable us to distinguish between nuclear (NSF) and magnetic (SF) scatterings.
Around the (1, 1, 0) reflection, at 2.3 K, $q_{\rm CDW}$ superlattice reflections in both $h$ and $k$ directions.
These contain both NSF and SF signals, as shown in Fig.~\ref{RS_prof}(b) and (c).
In contrast, the (1.58, 0, 0) reflection (Fig.~\ref{RS_prof}(d)) shows only the SF signal, indicating magnetic scattering without nuclear scattering contributions within the experimental accuracy.
Conversely, the (2, 0.42, 0) reflection (Fig.~\ref{RS_prof}(e)) presents a strong NSF signal but no detectable magnetic scattering.
This absence of magnetic signal is reasonable, since the condition of large values of $\alpha$ ($\sim 76^\circ$) and $\kappa$ makes it difficult to detect the $c$-plane transverse magnetic modulation.
The diffraction pattern of nuclear scatterings is consistent with that reported by Lee $et~al$\cite{Lee_2020}. as mentioned in the main text.
In conclusion, the polarized mode confirmed the extrinsic nature of the (i) magnetic signal in the (2, 0.42, 0) reflection and (ii) non-magnetic contribution in the (1.58, 0, 0) reflection, likely caused by multiple scatterings in the unpolarized mode.

\section{Calculations of magnetic scattering intensities}
Here, we present the formula of magnetic scattering intensities used in the quantitative analysis on the data obtained in the unpolarized mode.
The scattering intensity at scattering wave vector $\boldsymbol{\kappa}$ is related to the scattering cross section $\frac{d\sigma}{d\Omega}$ as follows:
\begin{equation}
I(\boldsymbol{\kappa})=KL(2\theta)\left(\frac{d\sigma}{d\Omega}\right).
\end{equation}
$K$ represents a scale factor, estimated to be $K = 21.4(5)$ through comparing observed intensities of nuclear Bragg reflections and the calculated scattering cross sections.
The cross sections of magnetic scattering is calculated in the following equation: 
\begin{equation}
\begin{aligned}
\frac{d\sigma}{d\Omega}&={\bm M}_{\perp}^*(\boldsymbol{\kappa}){\bm M}_{\perp}(\boldsymbol{\kappa}),\\
{\bm M}_{\perp}(\boldsymbol{\kappa})&\equiv \sum_{{\bm l}, {\bm d}}\left(-2.7 \boldsymbol{\mu}_{\perp}({\bm l}, {\bm d})f_{d}(\kappa) {\rm exp}(i\boldsymbol{\kappa}\cdot({\bm l}+{\bm d}))\right),
\end{aligned}
\end{equation}
where $\bm{l}$ and $\bm{d}$ denote the positions of a nuclear unit cell and the atomic positions in the nuclear unit cell, respectively.
For each magnetic reflection, ${\bm M}_{\perp}(\boldsymbol{\kappa})$ is as follows under the assumption that only U ions possess magnetic moments.

\vspace{1\baselineskip}
\noindent \textbf{(i) Magnetic reflections with ${\bm Q} = 0$}\\
In this case, two U ions in the nuclear unit cell have magnetic moments antiparallel to each other: $\boldsymbol{\mu}_{\perp}({\bm l}, {\bm d}) = \boldsymbol{\mu}_{\perp}$ (for U ion 1) and $-\boldsymbol{\mu}_{\perp}$ (for U ion 2).
Consequently, the cross section for this specific magnetic structure is determined as follows:
\begin{equation}
\begin{aligned}
\frac{d\sigma}{d\Omega}&=N\frac{(2\pi)^3}{v_0}|F_{m}(\boldsymbol{\kappa})|^2\delta(\boldsymbol{\kappa}-\boldsymbol{\tau}),\\
|F_{m}(\boldsymbol{\kappa})|&\equiv b_{mag}|\boldsymbol{\mu}_{\perp}|f_{\rm U}(\kappa)A(\boldsymbol{\kappa}).
\end{aligned}
\end{equation}
In this equation, $N$ represents the number of nuclear unit cells in the system, and $v_{0}$ is the volume of a nuclear unit cell.
The phase factor $A(\boldsymbol{\kappa})$ is defined and calculated using the equation:
\begin{equation}
A(\boldsymbol{\kappa})=\sum_{\bm l}{\rm exp}(i\boldsymbol{\kappa}\cdot{\bm d})=2.
\end{equation}

\noindent \textbf{(ii) Magnetic reflections with ${\bm q}_{\rm CDW}$}\\
The sinusoidal magnetic modulation in the $c$-plane is represented as follows:
\begin{equation}
\boldsymbol{\mu}_{\perp}({\bm l}, {\bm d}) = \boldsymbol{\delta\mu}_{\perp}\sin\left({\bm q}_{\rm CDW}\cdot({\bm l}+{\bm d})\right).
\end{equation}
The scattering cross section for this modulation is given by the following equation:
\begin{equation}
\begin{aligned}
\frac{d\sigma}{d\Omega}=N\frac{(2\pi)^3}{v_0}
  &\{|F^{+}_{m}(\boldsymbol{\kappa})|^2\delta(\boldsymbol{\kappa}+{\bm q}_{\rm CDW}-{\bm G})\\
  &+|F^{-}_{m'} (\boldsymbol{\kappa})|^2\delta(\boldsymbol{\kappa}-{\bm q}_{\rm CDW}-{\bm G})\},
\end{aligned}
\end{equation}
where ${\bm G}$ represents the reciprocal lattice vector, and $F^{\pm}_{m'}$ denotes the magnetic structure factor corresponding to the superlattice reflection at ${\bm G} \pm{\bm q}_{\rm CDW}$, which calculated as follows:
\begin{equation}
\begin{aligned}
|F^{\pm}_{m'}| &\equiv b_{mag}\frac{|\boldsymbol{\delta\mu}_{\perp}|}{2} f_{\rm U}(\kappa) A^\pm (\boldsymbol\kappa), \\
A^\pm (\boldsymbol\kappa) &= \sum_{\bm d}\exp\left(i(\boldsymbol{\kappa}\mp{\bm q}_{\rm CDW})\cdot {\bm d}\right).
\end{aligned}
\end{equation}
In this case, the phase factor $A^{\pm}(\boldsymbol{\kappa})$ is identical to that in the case (i),
thus, $A^{\pm}(\boldsymbol{\kappa}) = A(\boldsymbol{\kappa}) = 2$.

\vspace{1\baselineskip}
For both cases, the magnetic structure factors incorporate the factor of $N\frac{(2\pi)^3}{v_0}$, which is also present in the nuclear structure factor.
This factor is factored into the scale factor $K$ to calculate the magnitudes of $|F_{m}|$ and $|F^{\pm}_{m'}|$ by including this factor into the scale factor $K$ then obtain Eq. \eqref{Fmag_CDW} in the main text.
\bibliographystyle{jpsj}
\bibliography{UPt2Si2_neutron_jpsj_resubmittion_03}

\begin{verbatim}
\end{verbatim}
\end{document}